\documentclass[compsoc, conference, a4paper, 10pt, times]{IEEEtran}

\newcommand*{\SUBMISSION}{}%
\newcommand*{\CAMERAREADY}{}%

\newcommand{\sev}{\ac{SEV}}

\newcommand{\sevsnp}{\ac{SEV}-\ac{SNP}}
\newcommand{\snp}{\ac{SNP}}

\newcommand{\dmintegrity}{\texttt{dm-integrity}}
\newcommand{\dmverity}{\texttt{dm-verity}}
\newcommand{\dmcrypt}{\texttt{dm-crypt}}

\ifdefined\SUBMISSION
\usepackage[disable]{todonotes}
\else
\usepackage[textsize=footnotesize]{todonotes}
\presetkeys{todonotes}{fancyline}{}
\fi

\usepackage{acronym}
\usepackage{enumerate}
\usepackage{subcaption}
\usepackage{pgfplots}
\pgfplotsset{
    width=5cm, 
    legend style={
        font=\footnotesize,
    },
    compat=newest
}
\usepackage[utf8]{inputenc}
\DeclareUnicodeCharacter{2212}{−}
\usepgfplotslibrary{groupplots,dateplot}
\usetikzlibrary{patterns,shapes.arrows}
\usetikzlibrary{decorations.pathreplacing,calligraphy}

\usepackage{url}
\usepackage{tabularx}
\usepackage{fontawesome}
\usepackage{booktabs}
\usepackage{xspace}
\usepackage[pdftex,pdfauthor={Luca Wilke and Gianluca Scopelliti},pdftitle={SNPGuard}]{hyperref}

\usepackage[capitalize]{cleveref}
\crefname{figure}{Fig.\@}{Figs.\@}
\Crefname{figure}{Fig.\@}{Figs.\@}
\crefname{table}{Tab.\@}{Tabs.\@}
\Crefname{table}{Tab.\@}{Tabs.\@}
\crefname{section}{Sect.\@}{Sects.\@}
\Crefname{section}{Sect.\@}{Sects.\@}
\crefname{req}{Req.\@}{Reqs.\@}
\Crefname{req}{Req.\@}{Reqs.\@}

\usepackage{paralist}
\usepackage{crossreftools}

\usepackage{fix-cm} 

\DeclareMathAlphabet{\mathsfit}{T1}{\sfdefault}{\mddefault}{\sldefault}
\SetMathAlphabet{\mathsfit}{bold}{T1}{\sfdefault}{\bfdefault}{\sldefault}

\usepackage{listings}

\lstdefinestyle{mystyle}{
    inputpath=listings,
    commentstyle=\color{teal},
    keywordstyle=\color{magenta},
    numberstyle=\tiny\color{gray},
    stringstyle=\color{purple},
    basicstyle=\ttfamily\scriptsize,
    breakatwhitespace=false,         
    breaklines=true,                 
    captionpos=b,                    
    keepspaces=true,                 
    numbers=none,                    
    numbersep=5pt,                  
    showspaces=false,                
    showstringspaces=false,
    showtabs=false,                  
    tabsize=2
}

\acrodef{CC}{Confidential Computing}
\acrodef{CCA}{Confidential Compute Architecture}
\acrodef{CEK}{Chip Endorsement Key}
\acrodef{CRL}{Certificate Revocation List}
\acrodef{CSP}{Cloud Service Provider}
\acrodef{CVM}{Confidential Virtual Machine}
\acrodef{AWS}{Amazon Web Services}
\acrodef{ES}{Encrypted State}
\acrodef{GCP}{Google Cloud Platform}
\acrodef{IMA}{Integrity Measurements Architecture}
\acrodef{MitM}{Man-in-the-Middle}
\acrodef{PSP}{Platform Security Processor}
\acrodef{RMP}{Reverse Map Table}
\acrodef{RTMR}{Runtime Measurement Register}
\acrodef{SEV}{Secure Encrypted Virtualization}
\acrodef{SGX}{Security Guard Extensions}
\acrodef{SNP}{Secure Nested Paging}
\acrodef{SP}{Secure Processor}
\acrodef{SVSM}{Secure VM Service Module}
\acrodef{TCB}{Trusted Computing Base}
\acrodef{TDX}{Trust Domain Extensions}
\acrodef{TEE}{Trusted Execution Environment}
\acrodef{TPM}{Trusted Platform Module}
\acrodef{UKI}{Unified Kernel Image}
\acrodef{vTPM}{virtual TPM}
\acrodef{VCEK}{Versioned Chip Endorsement Key}
\acrodef{VLEK}{Versioned Loaded Endorsement Key}
\acrodef{VM}{Virtual Machine}
\acrodef{VMPL}{Virtual Machine Privilege Level}

\begin{document}

\title{SNPGuard: Remote Attestation of SEV-SNP VMs Using Open Source Tools}


\ifdefined\ANON
\author{\em Anonymous Author(s)}
\else

\ifdefined\CAMERAREADY
\makeatletter
\newcommand{\linebreakand}{%
  \end{@IEEEauthorhalign}
  \hfill\mbox{}\par
  \mbox{}\hfill\begin{@IEEEauthorhalign}
}
\makeatother

\author{\IEEEauthorblockN{Luca Wilke}
\IEEEauthorblockA{\textit{University of Lübeck, Germany} \\
l.wilke@uni-luebeck.de}
\and
\IEEEauthorblockN{Gianluca Scopelliti}
\IEEEauthorblockA{\textit{Ericsson Security Research, Sweden} \\
\textit{KU Leuven, Belgium} \\
gianluca.scopelliti@ericsson.com}
}
\else
\author{\IEEEauthorblockN{Luca Wilke$^\dagger{}$, Gianluca Scopelliti$^{*\ddagger{}}$}
\IEEEauthorblockA{
  \textit{$^{\dagger}$University of Lübeck, Germany
          $^{*}$Ericsson Security Research, Sweden;
          $^{\ddagger{}}$DistriNet, KU Leuven, Belgium} \\
  \textit{l.wilke@uni-luebeck.de; gianluca.scopelliti@ericsson.com}
}
}
\fi

\fi

\maketitle

\begin{abstract}
  Cloud computing is a ubiquitous solution to handle today's complex computing
demands. However, it comes with data privacy concerns, as the cloud service
provider has complete access to code and data running on their infrastructure.
VM-based Trusted Execution Environments (TEEs) are a promising solution to solve
this issue. They provide strong isolation guarantees to lock out the cloud
service provider, as well as an attestation mechanism to enable the end user to
verify their trustworthiness. Attesting the whole boot chain of a VM is a
challenging task that requires modifications to several software components.
While there are open source solutions for the individual components, the tooling
and documentation for properly integrating them remains scarce. In this paper,
we try to fill this gap by elaborating on two common boot workflows and
providing open source tooling to perform them with low manual effort. The first
workflow assumes that the VM image does only require integrity but not
confidentiality, allowing for an uninterrupted boot process. The second workflow
covers booting a VM with an encrypted root filesystem, requiring secure
provisioning of the decryption key during early boot. While our tooling targets
AMD Secure Encrypted Virtualization (SEV) VMs, the concepts also apply to other
VM-based TEEs such as Intel Trusted Domain Extensions (TDX).

\end{abstract}

\section{Introduction}

Cloud computing revolutionized the way businesses and individuals utilize
computing resources. Instead of owning and maintaining physical servers or data
centers, users rent them from large cloud service providers on a pay-as-you-go
basis, allowing unprecedented flexibility and scalability. One major concern
with shifting computation to the cloud is data privacy. \acp{TEE} aim to solve
this challenge, allowing to execute software on remote machines without the need
to trust their operator. To this end, they provide strong (cryptographic)
runtime isolation guarantees as well as an attestation feature to prove their
trustworthiness to the end user. Early designs like Intel \ac{SGX} were
process-scoped and required software to be rewritten, hindering adoption. More
recent \ac{TEE} designs like AMD \sev{} and Intel \ac{TDX} use a different
approach by protecting whole \acp{VM} to achieve a \textit{lift and shift}
solution. These \acp{VM} are also known as \acp{CVM}. Similar to booting a
physical machine, booting a \ac{CVM} includes several software components and
layers. As result, the correct attestation of \acp{CVM} is quite complex,
requiring changes to several components of the software stack such as guest
firmware and kernel. Nonetheless, without proper (remote) attestation the VM
owner cannot verify that their \ac{CVM} has not been tampered with. Cloud
providers like Microsoft Azure~\cite{azureSev} have solved this challenge, but
their solution relies on closed-source, proprietary software. As a result, the
\ac{VM} owner cannot verify these components, requiring them to again trust the
cloud service provider, partially defeating the purpose of \acp{TEE}.

In this paper, we introduce an easy-to-use, open source tool that tackles this challenge for
AMD \sevsnp{}, the latest iteration of \sev{}. While there already exist several individual components that are
open source, to the best of our knowledge there is no unified tool that provides
a full workflow from the initial \ac{CVM} preparation to its deployment,
attestation, and secret provisioning. As a result, a high manual effort and some
expertise are still required by the \ac{VM} owner. Our tool solves this issue by
integrating these existing components and providing templates for common use
cases.

In principle, our ideas should generalize to similar VM-based \ac{TEE}
technologies such as Intel \ac{TDX}, and in the future we aim at supporting them
as well. Our main goal is to provide a solid foundation for future research that
wants to either build on top of \acp{CVM} or experiment with tweaking
the individual components. We also hope that our tool will make \acp{CVM} more
accessible to developers. While our design is technically similar to
Revelio~\cite{revelio}, their code does not seem to be publicly available.
Additionally, in contrast to libkrun~\cite{libkrun}, which tries to minimize
\ac{VM} size and boot time, we use the feature-rich reference implementations
for the individual components. Finally, Pontes et
al.~\cite{DBLP:conf/ladc/PontesSFB23} developed an open source integration for
\sevsnp{} attestation with the SPIFFE framework~\cite{spiffee}. SPIFFE aims to
standardize secret injection into services. In contrast to our tool, Pontes et
al. do not provide tooling for configuring and setting up the \sevsnp{} VM. In
addition, their dependency on SPIFFE requires a more complex setup, while our tool is focussed on providing building
blocks that can be easily set up and customized.

The remainder of this paper is organized as follows: \cref{chapter:background}
defines the attacker scenario and gives background on AMD \sevsnp{}. In
\cref{chapter:design} and \cref{chapter:implementation} we describe our tool and
give implementation details. In \cref{chapter:discussion} we discuss remaining
problems and future work, and in \cref{chapter:conclusions} we summarize our
paper.

\section{AMD SEV-SNP in a Nutshell}\label{chapter:background}

\sevsnp{}~\cite{AmdSevSnpWhitepaper,AmdSevAbi} is a \ac{TEE} that protects the
confidentiality and integrity of whole \acp{VM} against an attacker with root
privileges and physical access to the machine, enabling to run SEV-protected
\acp{VM} without trusting the infrastructure provider and virtualization layers
such as the hypervisor.

The root-of-trust of an AMD CPU is the AMD \ac{SP} (formerly known as PSP), a
coprocessor that hosts the \sevsnp{} firmware and performs security-critical
operations like generating and managing memory encryption keys. Each protected
\ac{VM} is assigned a unique AES-128 key that is used to encrypt its data before
it is written to the RAM. Inside the CPU, an access-right based system is used
to protect data in the guest \ac{VM}. In addition, the \ac{RMP} prevents the
hypervisor or other tenants from writing to \ac{VM} memory, ensuring integrity.
Moreover, \sevsnp{} introduced \acp{VMPL}, an additional set of hardware
enforced privilege levels than can  be used by the \ac{VM}. This enables use
cases like running a \ac{vTPM} inside a secure \ac{VM}, on a higher privilege
layer than the rest of the software stack.

The AMD \ac{SP} contains a chip-unique signing key called \ac{CEK}, derived from
secrets stored in chip's fuses. This key is used to derive other keys such as
the \ac{VCEK}, used to sign attestation reports. The \ac{VCEK} also depends on
the version number of the \ac{SP} firmware and the CPU microcode. This allows a
verifier to attest that the platform where a secure \ac{VM} is running has an
up-to-date software stack. A public key infrastructure is used to certify the
\ac{VCEK} from AMD's root certificates, ensuring the authenticity of attestation
reports. Recently, the \ac{SP} firmware started to support an alternative to the
\ac{VCEK} called \ac{VLEK}, which identifies a specific cloud provider. When
requesting an attestation report, the \ac{VCEK} and \ac{VLEK} can be used
interchangeably.

\section{Design}\label{chapter:design} In this section we describe the high
level design of our tool. Implementation details will be given in
\cref{chapter:implementation}.

We designed our tool with two main goals in mind:
\begin{enumerate}
    \item \emph{Integrity of the VM boot chain.} It is paramount that we ensure
    that the deployed \ac{VM} is in a good, known state. This is typically done
    via remote attestation, although confidential \acp{VM} have additional
    challenges because their boot process follows multiple stages. Normally, the
    launch measurement included in the \sevsnp{} attestation report only covers
    the guest firmware, hence there is a need to extend this measurement to
    later stages such as kernel, initial RAM disk (\emph{initramfs}), and
    eventually the root filesystem.
    \item \emph{Secret provisioning.} A typical confidential workflow involves
    injecting some kind of secrets into the deployed \ac{TEE} instance, e.g.,
    private keys, sensitive data such as personal identifiable information, or
    intellectual property such as AI models. Our tool supports two modes of
    secret provisioning, i.e., at runtime and at rest. In the first case, the VM
    initially does not contain any secrets, which are only injected after boot
    via a secure channel established upon remote attestation. In the second
    case, the VM's root filesystem requires confidentiality and is thus
    encrypted before transmission to the hypervisor. As a result, we need to
    enable provisioning the disk encryption key during early boot.
\end{enumerate}

\begin{figure}
    \centering
    \includegraphics[width=0.45\textwidth]{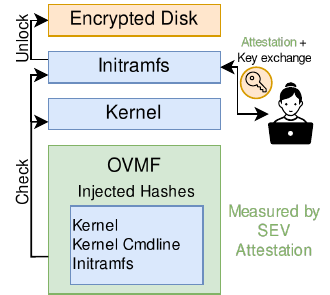}
    \caption{Trust chain of the software stack running inside the VM that shows how to use the SEV measurement as a trust anchor for software-based measurements. The control flow is bottom to top.}
    \label{figure:trust-chain}
\end{figure} 
To achieve the above goals, our tool uses a two-stage approach. The first stage
covers the first part of the \ac{VM} boot process and includes guest firmware,
kernel, initramfs and kernel command line parameters. Here, we assume that these
components do not contain any secrets and come from known (e.g., open source)
code. Thus, we only need to ensure their integrity (Goal 1). The second stage
consists of the \ac{VM}'s root filesystem, which may or may not contain secrets
(Goal 2). Here, integrity is always mandatory, but confidentiality is optional.
To ensure the integrity of the first stage, we use the \sevsnp{} isolation and
attestation features. To achieve integrity and confidentiality for the second
stage, we use the disk protection techniques supported by the Linux kernel.

\subsection{First Stage}\label{design:first-stage}

The very first piece of code that is executed after starting the \sevsnp{}
\ac{VM} is the firmware, which has to be \sevsnp{}-aware in order to correctly
set up the \ac{VM}, e.g., by allocating memory pages and configuring page bits.
To achieve this, we use the \sevsnp-enabled version of the
OVMF~\cite{amdsevGithub} UEFI implementation.

Normally, the launch measurement in the \sevsnp{} attestation report only
includes the memory content of the OVMF firmware and the initial register state
of the \ac{VM}'s virtual CPUs. However, there is an \snp{}-enabled OVMF firmware
version that supports extending the measurement to include the kernel, the
initramfs and the kernel command-line parameters, which are then loaded using
Direct Linux Boot~\cite{directLinuxBoot}. \cref{figure:trust-chain} shows an
overview of the mechanism. In short, before loading OVMF into memory, the
hypervisor measures those components and adds the corresponding hashes in a
special section of the OVMF binary. Afterwards, the ``hot-patched'' version of
OVMF is loaded into memory, making the injected hash values for the kernel,
initramfs and kernel command line part of the \sevsnp{} launch measurement. At
runtime, before transferring the execution to the Linux kernel, OVMF re-computes
the measurements of each component and compares them against the stored
reference value, failing to boot if they do not match. As a result, we can use
the remote attestation features of \sevsnp{} to prove the integrity and
authenticity of the first stage to the \ac{VM} owner. 

With \sevsnp, the code inside the \ac{VM} is responsible for requesting the
attestation report. The report is signed by the AMD \ac{SP} whose public key can
be verified using a publicly available certificate chain. Thus, it is safe to
transmit the attestation report over an unprotected channel. To prevent an
attacker from replaying an old attestation report, we make use of the ``Guest
Data'' field, which allows the \ac{VM} to include 64 bytes of arbitrary data
into the signed attestation report, to include a nonce sent by the \ac{VM}
owner. By checking that the report contains the expected nonce, the VM owner can
verify that they received a fresh report.

\subsection{Second Stage}\label{design:second-stage}

With the first stage, we can ensure the integrity of the guest \ac{VM} up to the
so-called \emph{early userspace}, i.e., where the initial RAM disk is mounted
into memory and some initial configuration of the \ac{VM} is done before passing
control to the ``real'' root filesystem. However, as \sevsnp{} does not provide
protection of code and data at rest, we need other mechanisms to ensure the
integrity and (optionally) confidentiality of persistent disks, which can
otherwise be easily accessed by the hypervisor.

Depending on whether the root filesystem requires confidentiality, our tool
offers two different approaches to secret provisioning (Goal 2). If the root
filesystem does not contain any secret data, the \ac{VM} can boot uninterrupted
and we can later verify that the \ac{VM} is in a good state via attestation.
Secrets may still be injected after the \ac{VM} has fully booted. 
If the root filesystem should remain secret, we need to provision it in
encrypted form. As a result, the \ac{VM} cannot boot past the first stage without having provisioned
the decryption key. Therefore, we necessarily need to perform both attestation
and secret provisioning \emph{during} early boot. Below, we describe the two
approaches in more detail.

\noindent\textbf{\emph{Integrity-only Workflow}}
The main idea here is that we do not want to ``pause'' the boot process, but
instead allow the \ac{VM} to boot without any intervention, and only later,
whenever needed, attest it and provision secrets. To achieve this goal, we
leverage the \dmverity{} utility~\cite{dmVerity} to ensure the integrity of the
root filesystem. In short, the filesystem is divided into blocks of a predefined
size, and each of them is measured using a hash function (e.g., SHA-256). All
hashes are then stored in a Merkle tree on a separate data disk, leaving the
original root filesystem unmodified. The root of the Merkle tree, called
\emph{root hash}, guarantees the integrity of the whole filesystem and needs to
be stored separately. When the root filesystem is mounted, during early
userspace, \dmverity{} takes as input the data disk and root hash and verifies
the integrity the filesystem. Besides, integrity is checked at each read,
resulting in a kernel panic if a violation is discovered.

With this approach, the integrity of the root filesystem can be reduced to the
integrity of the root hash, which has to be provisioned to the guest \ac{VM}
without tampering, but does not require confidentiality. We solve this problem
by adding the hash as a kernel parameter. As the first stage already ensures the
integrity of the kernel command-line parameters, this approach naturally extends the
\ac{VM}'s integrity guarantees to the root filesystem. This can be verified, at
any moment, by the usual remote attestation procedure.

One caveat of \dmverity{} is that the root filesystem is mounted as read-only,
which may not be acceptable for all applications. Besides, there are partitions
in the filesystem that need to be read-write for the VM to function properly,
e.g., \texttt{/home}, \texttt{/tmp} and \texttt{/var}. We solve this problem by
creating one \texttt{tmpfs} filesystem~\cite{tmpfs} for each partition that
needs to be read-write. These filesystems will only reside in memory, thus
their integrity and confidentiality are preserved at runtime by \sevsnp{}. This
process is done in the initramfs where, after opening and verifying the
\dmverity{} filesystem, files are copied to the \texttt{tmpfs} partitions.

Clearly, secrets such as cryptographic keys should be deleted from the root
filesystem when choosing this workflow, as there is no confidentiality at rest.
Our tool helps with this process by automatically deleting all SSH keys from
\texttt{/etc/ssh} when preparing the \dmverity{} filesystem. Those keys are then
regenerated in early userspace and stored in the \texttt{tmpfs} partitions, such
that the guest owner can safely connect to the \ac{CVM} at runtime. Afterwards,
they can retrieve the attestation report to verify its boot chain.

\noindent\textbf{\emph{Integrity+Confidentiality Workflow}}
\begin{figure}
    \centering
    \includegraphics[width=0.45\textwidth]{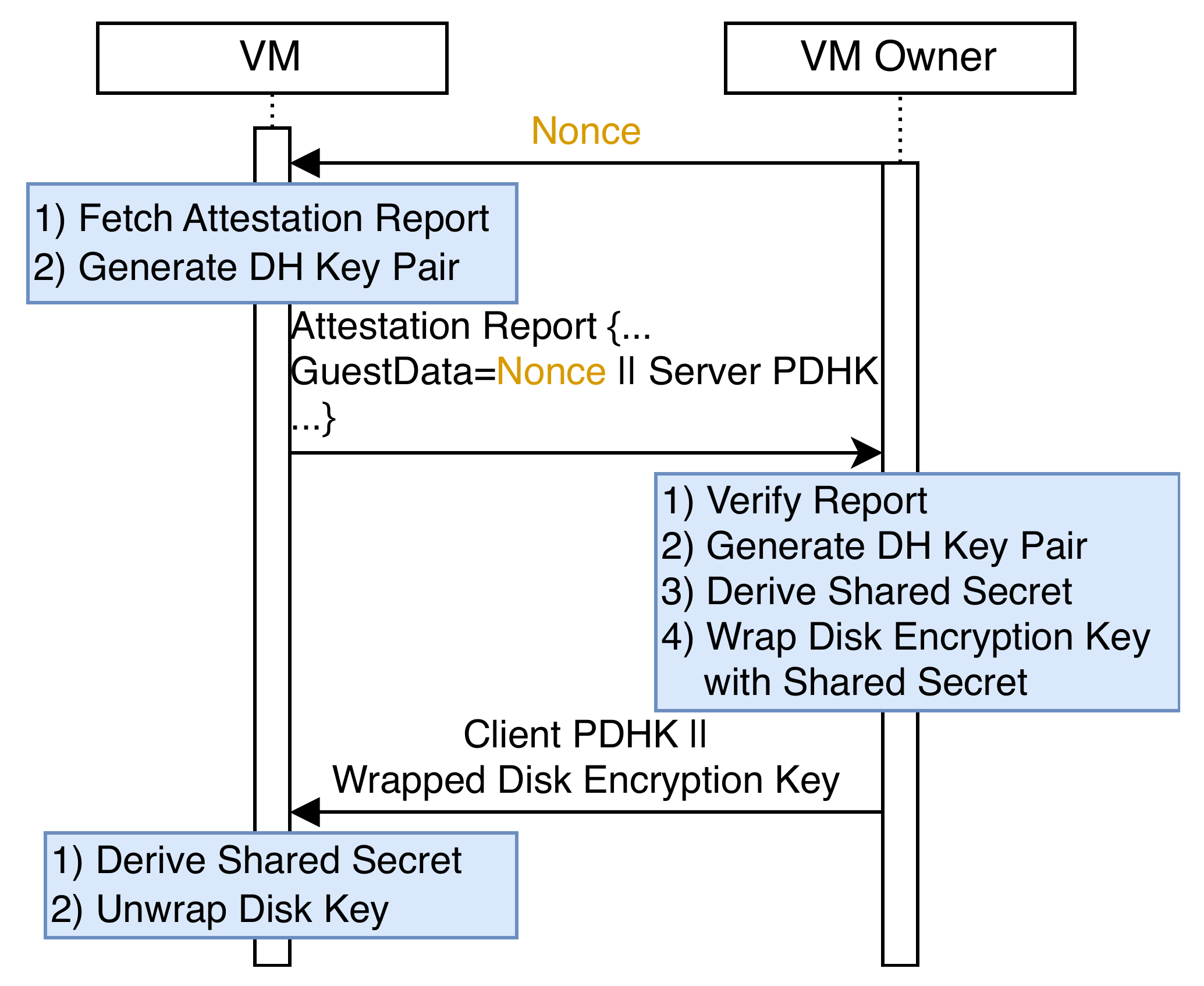}
    \caption{Protocol that uses the Guest Data field in the attestation report to securely send a disk encryption key to the VM. DH stands for Diffie-Hellman and PDHK refers to the public part of the DH key.}
    \label{figure:attestation-protocoll}
\end{figure} 
For this workflow, the \ac{VM} owner encrypts the root filesystem with
\dmcrypt~\cite{dmCrypt} before sending it to the untrusted hypervisor.
\dmcrypt{} can also be configured to ensure integrity protection by leveraging
\dmintegrity~\cite{dmIntegrity} under the hood. To unlock the encrypted disk for
the second stage, the first stage needs access to the encryption key. Thus, we
need to build an encrypted channel between the \ac{VM} owner and the \ac{VM} to
transfer the key. To achieve this, we again leverage the ``Guest Data'' field to
perform a Diffie-Hellman (DH) key exchange, as depicted
in~\Cref{figure:attestation-protocoll}. The \ac{VM} owner starts the process by
requesting the attestation report, using a nonce to ensure freshness. Next, the
\ac{VM} generates an ephemeral DH key and includes the public part in the
``Guest Data'' field, together with the nonce. After verifying the attestation
report, the \ac{VM} owner uses their ephemeral DH key pair to derive a shared
secret and uses it to encrypt the disk encryption key using an authenticated
encryption scheme. Finally, the \ac{VM} owner sends both the encrypted disk
encryption key and the public part of their DH key to the \ac{VM}. While the
\ac{VM} authenticates to the guest owner by providing a valid attestation
report, the \ac{VM} owner implicitly proves their authenticity to the \ac{VM} by
providing the correct disk encryption key.

The code inside the \ac{VM} uses the public DH key of the \ac{VM} owner to
derive the shared secret as well, allowing it to decrypt the disk encryption
key. Now, we can unlock and transfer control to the root filesystem by executing
its \textit{init} script, bringing up the rich userspace environment of the
\ac{VM}.
After the initialization of the second stage has completed, the \ac{VM} owner
can, e.g., use SSH to connect to the \ac{VM}. Since the second stage image
was encrypted before it was transferred to the untrusted hypervisor, the \ac{VM}
owner can securely store the private key of the SSH server as well as the list
of public keys that are allowed to log in inside the image.

\noindent\textbf{\emph{Comparison}}
Both workflows have pros and cons. While the integrity-only workflow does not
offer confidentiality at rest, the encrypted workflow necessarily requires users
to pause the boot process to get the root filesystem's decryption key. Besides,
\dmverity{} might have a better I/O performance compared to
\dmcrypt{}+\dmintegrity, but at the same time mounting some partitions as
\texttt{tmpfs} in the integrity-only workflow would incur some boot latency (for
copying files) and increase memory usage. Users should therefore choose the
workflow that is best suitable for their use case.

\section{Implementation}\label{chapter:implementation}

This section gives implementation details for the design outlined in
\cref{chapter:design}. For the hypervisor, we use the patched Linux kernel and
QEMU versions provided by AMD~\cite{amdsevGithub} in order to run
\sevsnp{}-enabled \acp{VM}. For the \ac{VM}, we use the \sevsnp{} target of OVMF~\cite{OVMF} as the UEFI, and a self-written client and server for the attestation workflow.
In addition, we developed scripts to streamline the whole setup process.
Our tool is available at
\ifdefined\ANON
\url{https://github.com/anonymized}.
\else
\url{https://github.com/SNPGuard/snp-guard}.
\fi

\noindent\textbf{\emph{Building the first stage image.}}
Our first stage image consists of the OVMF UEFI code plus a Linux kernel, its
command-line parameters, and an initramfs. The latter three can be provided
either separately or as a \ac{UKI}. Both OVMF and the Linux kernel need to be
adapted to work with \sevsnp. Thus, we use AMD's official
forks~\cite{amdsevGithub} for both. However, most features are also available in
the official upstream versions. For OVMF, we need to build the
\textit{AmdSev/AmdSevX64.dsc} target instead of the default
\textit{OvmfPkgX64.dsc} target used by the build script in the AMD repository.
This enables OMVF's support for extending the \sev{} launch measurement to the
kernel (cf. \cref{chapter:design}).

\noindent\textbf{\emph{Building the initramfs.}}
We require modifications to the initramfs to support integrity and encryption of
the root filesystem. The initramfs is a minimal filesystem that contains
scripts, kernel modules, applications and configuration files that are required
to bootstrap the VM and mount the root filesystem.

To build our custom initramfs, we leverage Docker~\cite{docker} to create a
container image with all the required tools. This approach allows us to easily
install components with complex dependencies. Afterwards, we export all layers
of our container image into a single folder and convert it to the initramfs
format. For the \textit{init} binary inside the initramfs we use a small shell
script that calls a series of applications to orchestrate the boot process and
eventually mounts the filesystem for the second stage. Next, the \textit{init}
script uses \textit{switch\_root}~\cite{switchRoot} to unmount the current
initramfs filesystem and jump into the root filesystem using the contained
\textit{init} script as the entry point for the execution, taking care of
starting the services contained in the second stage. One caveat with this
approach is that the second stage will still use the initially booted Linux
kernel. We discuss implications and potential solutions
in~\cref{chapter:discussion}.

If full disk encryption is used, the first stage needs access to the disk
encryption key before it can proceed with the boot process. To provision the
key, we perform the remote attestation protocol described
in~\cref{design:second-stage} which we implemented based on the open source
\textit{sev} library developed by the VirTEE community~\cite{virtee}.

\noindent\textbf{\emph{Building the second stage image.}}
For preparing the root filesystem that is mounted in the second stage we start
from an existing Linux filesystem that can be created locally via usual means,
such as creating a new \ac{VM} from scratch and installing a fresh Linux
distribution. Afterwards, we provide tools and scripts to extract the filesystem
from the \ac{VM} image, make the necessary adjustments, and generate the
corresponding Merkle tree (if \dmverity{} is used) or to create an encrypted
copy (if \dmcrypt{} is used).

\section{Limitations and Future Work}\label{chapter:discussion}

\noindent\textbf{\emph{Flexible boot.}}
One limitation of our \textit{switch\_root}-based approach for transitioning
from the first stage to the second stage is that it does not change the kernel
itself. Thus, we are still executing the publicly known kernel from the first
stage. Furthermore, any changes that the second stage \ac{VM} does to the kernel
or initramfs stored on its \textit{/boot} partition do not have any effect. One
potential solution for this issue is to use the Linux kernel's
\textit{kexec}~\cite{kexec} feature, which enables booting into a new Linux
kernel from an already running kernel. However, we were not able to use this
feature as \sevsnp{} support for \textit{kexec} seems to be work in
progress~\cite{kexecForSev}.

A more principled solution would be to change the first stage image to only
include a bootloader like Grub or systemd-boot, instead of a full Linux kernel.
This way, we could properly boot the second stage using its kernel and
initramfs. In addition, this decouples the launch measurement from the second
stage image as we no longer need to change the first stage if we want to use a
different kernel version with our second stage. The main drawback of this
approach is that it would result in a quite restricted programming environment
for the first stage, which conflicts with our goal of enabling an easy
integration of experimental changes. We leave the implementation of this
approach for a future version of our tool.

\noindent\textbf{\emph{Cloud deployments.}}
Currently, our tool can be used to deploy \acp{CVM} on \sevsnp{}-enabled hosts
where the guest owner can control the boot process and the parameters passed to
the \ac{CVM} at launch. However, cloud providers such as
\ac{AWS}~\cite{awsSevSnp}, Microsoft Azure~\cite{azureSev}, and
\ac{GCP}~\cite{googleConfidentialComputing} currently only provide limited
customization options in their confidential computing offering where, e.g., it
is not possible to run a custom guest firmware. Unfortunately, this means that
our workflows are not fully supported on public clouds. While we hope that this
situation will change in the future, users might still benefit from our tool at
present by building individual components (e.g., encrypting the root
filesystem).

\noindent\textbf{\emph{Virtual TPM.}}
In the future, we are planning to support a \ac{vTPM} inside the guest
\ac{VM}\@. In \sevsnp{}, this can be achieved by leveraging the \sevsnp{}
specific \ac{VMPL} (c.f.~\cref{chapter:background}) feature to run a module at the
highest VMPL privilege level that exposes \ac{vTPM} functionalities to lower-privileged
levels such as the guest OS\@. Besides, the \ac{vTPM} can also be used to
measure the first stage of boot without requiring to inject kernel measurements
into OVMF (cf.~\cref{design:first-stage}). Moreover, a \ac{vTPM} can be combined
with the Linux \ac{IMA}~\cite{sailer2004ima} to provide runtime attestation of
userspace applications, which would enable integrity protection of the second
stage with minimal changes (cf.~\cref{design:second-stage}). The open source
community is actively working on \ac{vTPM} support as part of the \ac{SVSM}
project~\cite{coconutSVSM}, and a full design has been proposed by Narayanan et
al.~\cite{narayanan2023svsm}.

\noindent\textbf{\emph{Intel TDX.}}
The design in \cref{chapter:design} does not only apply to \sevsnp{} but can be
generalized to all VM-based \acp{TEE}. In particular, it would be worthwhile to
extend our tool to also support Intel \ac{TDX}. Here, the only difference would
concern the first stage to build \ac{TDX}-aware OVMF and kernel binaries.
Besides, we could leverage \ac{TDX}'s \acp{RTMR}~\cite{intelTdxModule} to
perform integrity measurements similarly to a \ac{vTPM} (possibly combined with
Linux \ac{IMA}).

\section{Conclusions}\label{chapter:conclusions}

Recent advancements in the confidential computing landscape have highlighted
that there is a need to ease the development of \ac{TEE}-based solutions in
order to increase adoption, making this technology accessible even to
non-skilled developers. While \ac{VM}-based \acp{TEE} are a significant step in
the right direction as they do not require any modifications in user
applications, a lot of manual effort is still required to configure and properly
attest \ac{CVM} workloads. With our work, we try to further lower the bar
towards \ac{TEE} adoption by minimizing the manual effort required to provision
and attest \acp{CVM}, and we hope that our tool can accelerate future \ac{TEE}
research or even be integrated into the supply chain of cloud workloads in the
future.

\ifdefined\CAMERAREADY
\subsection*{Acknowledgements}
This research is partially funded by the Research Fund KU Leuven, by the
Cybersecurity Research Program Flanders, by EU H2020 MSCA-ITN action 5GhOSTS under grant agreement no. 814035 and by the BMBF project SASVI.
\fi

\bibliographystyle{IEEEtranS}
\bibliography{bibliography}

\begin{thebibliography}{10}
\providecommand{\url}[1]{#1}
\csname url@samestyle\endcsname
\providecommand{\newblock}{\relax}
\providecommand{\bibinfo}[2]{#2}
\providecommand{\BIBentrySTDinterwordspacing}{\spaceskip=0pt\relax}
\providecommand{\BIBentryALTinterwordstretchfactor}{4}
\providecommand{\BIBentryALTinterwordspacing}{\spaceskip=\fontdimen2\font plus
\BIBentryALTinterwordstretchfactor\fontdimen3\font minus
  \fontdimen4\font\relax}
\providecommand{\BIBforeignlanguage}[2]{{%
\expandafter\ifx\csname l@#1\endcsname\relax
\typeout{** WARNING: IEEEtranS.bst: No hyphenation pattern has been}%
\typeout{** loaded for the language `#1'. Using the pattern for}%
\typeout{** the default language instead.}%
\else
\language=\csname l@#1\endcsname
\fi
#2}}
\providecommand{\BIBdecl}{\relax}
\BIBdecl

\bibitem{amdsevGithub}
{AMD}, ``{AMDSEV Github Repository},''
  \url{https://github.com/AMDESE/AMDSEV/tree/snp-latest}, {Accessed on March
  23, 2024}.

\bibitem{AmdSevSnpWhitepaper}
------, ``{AMD SEV-SNP: Strengthening VM Isolation with Integrity Protection
  and More},'' {White Paper}, January 2020.

\bibitem{AmdSevAbi}
------, ``{SEV Secure Nested Paging Firmware ABI Specification},''
  {Specification}, September 2023, version 1.55.

\bibitem{awsSevSnp}
{AWS}, ``{AMD SEV-SNP},''
  \url{https://docs.aws.amazon.com/AWSEC2/latest/UserGuide/sev-snp.html}.

\bibitem{coconutSVSM}
{COCONUT}, ``{COCONUT-SVSM Github Repository},''
  \url{https://github.com/coconut-svsm/svsm}, {Accessed on March 23, 2024}.

\bibitem{docker}
{Docker}, ``{Docker},'' \url{https://www.docker.com/}, {Accessed on March 26,
  2024}.

\bibitem{revelio}
\BIBentryALTinterwordspacing
A.~Galanou, K.~Bindlish, L.~Preibsch, Y.-A. Pignolet, C.~Fetzer, and
  R.~Kapitza, ``Trustworthy confidential virtual machines for the masses,'' in
  \emph{Proceedings of the 24th International Middleware Conference}, ser.
  Middleware '23.\hskip 1em plus 0.5em minus 0.4em\relax New York, NY, USA:
  Association for Computing Machinery, 2023, p. 316–328. [Online]. Available:
  \url{https://doi.org/10.1145/3590140.3629124}
\BIBentrySTDinterwordspacing

\bibitem{googleConfidentialComputing}
{Google Cloud}, ``{Confidential Computing},''
  \url{https://cloud.google.com/security/products/confidential-computing}.

\bibitem{intelTdxModule}
{Intel}, ``{Intel Trust Domain Extensions (Intel TDX) Module Base Architecture
  Specification},'' {Specification}, November 2023, version 348549-003US.

\bibitem{switchRoot}
P.~Jones, J.~Katz, and Z.~Karel, ``switch\_root man page,''
  \url{https://man7.org/linux/man-pages/man8/switch_root.8.html}, {Accessed on
  March 22, 2024}.

\bibitem{kexecForSev}
V.~Karasulli, ``x86/snp: Add kexec support,''
  \url{https://lwn.net/Articles/965773/}, {Accessed on March 22, 2024}.

\bibitem{dmCrypt}
{kernel.org}, ``dm-crypt,''
  \url{https://www.kernel.org/doc/html/latest/admin-guide/device-mapper/dm-crypt.html},
  {Accessed on March 22, 2024}.

\bibitem{dmIntegrity}
------, ``dm-integrity,''
  \url{https://docs.kernel.org/admin-guide/device-mapper/dm-integrity.html},
  {Accessed on March 22, 2024}.

\bibitem{dmVerity}
------, ``dm-verity,''
  \url{https://docs.kernel.org/admin-guide/device-mapper/verity.html},
  {Accessed on March 22, 2024}.

\bibitem{tmpfs}
------, ``Tmpfs file system,''
  \url{https://www.kernel.org/doc/html/latest/filesystems/tmpfs.html},
  {Accessed on May 14, 2024}.

\bibitem{kexec}
{Linux man pages}, ``kexec man page,'' \url{https://linux.die.net/man/8/kexec},
  {Accessed on March 22, 2024}.

\bibitem{libkrun}
S.~López, ``{libkrun Github Repository},''
  \url{https://github.com/containers/libkrun}.

\bibitem{azureSev}
{Microsoft Azure}, ``{About Azure confidential VMs},''
  \url{https://learn.microsoft.com/en-us/azure/confidential-computing/confidential-vm-overview}.

\bibitem{narayanan2023svsm}
\BIBentryALTinterwordspacing
V.~Narayanan, C.~Carvalho, A.~Ruocco, G.~Almasi, J.~Bottomley, M.~Ye,
  T.~Feldman-Fitzthum, D.~Buono, H.~Franke, and A.~Burtsev, ``{Remote
  attestation of confidential VMs using ephemeral vTPMs},'' in
  \emph{Proceedings of the 39th Annual Computer Security Applications
  Conference}, ser. ACSAC '23.\hskip 1em plus 0.5em minus 0.4em\relax New York,
  NY, USA: Association for Computing Machinery, 2023, p. 732–743. [Online].
  Available: \url{https://doi.org/10.1145/3627106.3627112}
\BIBentrySTDinterwordspacing

\bibitem{DBLP:conf/ladc/PontesSFB23}
\BIBentryALTinterwordspacing
D.~Pontes, F.~Silva, E.~D.~L. Falc{\~{a}}o, and A.~Brito, ``Attesting {AMD}
  {SEV-SNP} virtual machines with {SPIRE},'' in \emph{12th Latin-American
  Symposium on Dependable and Secure Computing, {LADC} 2023, La Paz, Bolivia,
  October 16-18, 2023}.\hskip 1em plus 0.5em minus 0.4em\relax {ACM}, 2023, pp.
  1--10. [Online]. Available: \url{https://doi.org/10.1145/3615366.3615419}
\BIBentrySTDinterwordspacing

\bibitem{directLinuxBoot}
{QEMU}, ``{Direct Linux Boot},''
  \url{https://qemu-project.gitlab.io/qemu/system/linuxboot.html}, {Accessed on
  March 26, 2024}.

\bibitem{sailer2004ima}
R.~Sailer, X.~Zhang, T.~Jaeger, and L.~Van~Doorn, ``{Design and implementation
  of a TCG-based integrity measurement architecture.}'' in \emph{USENIX
  Security symposium}, vol.~13, no. 2004, 2004, pp. 223--238.

\bibitem{spiffee}
{SPIFFE}, ``{Secure Production Identity Framework For Everyone (SPIFFE)},''
  \url{https://github.com/spiffe/spiffe}, {Accessed on May 7th, 2024}.

\bibitem{OVMF}
{tianocore}, ``{Open Virtual Machine Firmware (OVMF)},''
  \url{https://github.com/tianocore/edk2}, {Accessed on May 7th, 2024}.

\bibitem{virtee}
{VirTEE}, ``{VirTEE}: {A} community for building virt-based {TEEs},''
  \url{https://virtee.io/}, {Accessed on March 23, 2024}.

\end{thebibliography}


\end{document}